\documentclass[pre,preprint, groupedaddress, showkeys,showpacs]{revtex4}
\usepackage{graphicx}
\begin{document}

%\preprint{}

\title{The effects of environmental disturbances on tumor growth}


\author{Ning Xing Wang$^1$, Xiao Miao Zhang$^1$, Xiao Bing Han$^2$}
%\homepage[]{}

%\thanks{}
\affiliation{$^1$School of Physics and Telecommunication
Engineering, South China Normal University, Guangzhou, China\\
$^2$Guangdong No.2 Provincial People's Hospital, Guangzhou, China}


\date{\today}
\begin{abstract}
\indent In this study, the analytic expressions of the steady
probability distribution of tumor cells were established based on
the steady state solution to the corresponding Fokker-Planck
equation. Then, the effects of two uncorrelated white noises on
tumor cell growth were investigated. It was found that the predation
rate plays the main role in determining whether or not the noise is
favorable for tumor growth.
\end{abstract}



\maketitle
\section {Introduction}
As tumors seriously threaten human health, extensive attention has
been paid to this issue by researchers in various
fields\cite{1,2,3,4,5,6,7,8,9,10,11,12,13,14,15,16,17,18,19,20,21,22,23,24,25,26,27}.
It is known that tumor cell growth is a complex process, and is
governed by environmental fluctuations such as the people's
spiritual status, as well as other diseases from which they suffer.
Recently, researchers from the field of nonlinear physics have
introduced noise into the model of tumor cell growth, where $noise$
refers to the various disturbances involved in tumor growth. For
example, Ai and coworkers\cite{2,3,5} studied the effects of
correlated Gaussian white noise in a logistic growth model. This
model is often as a basic model for cell growth, particularly tumor
cell growth\cite{28,29}, to describe such growth under ideal
conditions without fluctuation. Ai and coworkers found that noise
during tumor cell growth can induce phase transition, and that
intensive environmental fluctuations may even cause the extinction
of tumor cells. Furthermore, Zhong and coworkers\cite{7,26}
investigated the random resonance of tumor growth with noise. It was
found that the steady distribution probability of tumor growth
changed from a uni-peak state to a bi-peak state when the intensity
of multiplicative noise increased. An appropriate intensity of
multiplicative noise can destroy the mechanism of tumor growth. In
contrast, superfluous noise can be beneficial for their growth. Mei
and coworkers\cite{9} investigated the tumor cell growth model in
the presence of correlated noises and found that the correlation
intensity $\lambda$ and correlation time $T$ play opposite roles in
the static properties and the state transition of the system.
 An increase in $\lambda$ can produce a smaller mean value of the cell population and slow down the state transition.
 However, an increase in $T$ can produce a larger mean value of the cell population and enhance the state transition.\\
The abovementioned results demonstrate that tumor growth models with
noise are closer to the real situation, although most models exhibit
some differences from the real process of tumor growth. In such
studies, researchers have attempted to obtain deeper insights into
the intrinsic mechanisms of tumor growth
and provide new ideas for tumor treatment.\\
The noises introduced into the tumor growth model are generally a
single noise, correlated multiplicative noise\cite{3,5}, etc. It has
been shown that different types and numbers of noise may be
operating in the context of tumor growth\cite{2,3,5,8,9,10,26}, and
the corresponding tumor growth behaviors have also been observed to
be different. The random fluctuations introduced in our study are
different from those in previous studies. In this paper, the effects
of two uncorrelated Gaussian white noises on tumor growth will be
studied. These are the effect of additive noise on the birth rate of
tumor cells and that of multiplicative noise on the predation rate
of anticancer cells. It is shown that the effect of noise on the
tumor growth is mainly determined by the predation rate of
anticancer cells. With changes in the parameters, the steady
distribution probability of tumor growth changes
between a single steady state and bi-stable state.\\

\section{THE DETERMINISTIC MODEL OF TUMOR CELL GROWTH}
Lefever and Garay\cite{30} studied tumor growth under immune
surveillance against cancer using the enzyme dynamics model. The
model is as follows:
\begin{center}{Normal Cells $\longrightarrow$ $X$,}\end{center}
\begin{center}{$X$ $\longrightarrow$ $2X$,}\end{center}
 \begin{center}{$X$ $+$ $E_{0}$ $\longrightarrow$ $E$ $\longrightarrow$ $E_{0}$ + $P$, }\end{center}
 \begin{center}{$P$ $\longrightarrow$,}\end{center}
Here, $X$, $P$, $E_{0}$, and $E$ are cancer cells, dead cancer
cells, immune cells, and the compounds of cancer cells and immune
cells, respectively. This model reveals that normal cells can
transform into cancer cells, and then the cancer cells reproduce,
decline, and ultimately die out.\\
Based on the model by Lefever and Garay, we investigate the Logistic
model of Verhulst, and only consider the growth of tumor cells and
anticancer cells. We assume that tumor cells satisfy the following
equation[31,32]:
\begin{equation}
\frac{dX}{dt}=r_{B}X\left(1-\frac{X}{k_{B}}\right)-P(X),
\end{equation}
where $X$ is the relative number of tumor cells, $r_{B}$ is the
birth rate of the tumor cells, and $k_{B}$ is carrying capacity.
$P(X)$ represents predation generated by anticancer cells. We take
the $P(X)$ expression suggested by Ludwing[31]. In his work $P(X)$
is expressed by $BX^{2}/(A^{2}+X^{2})$, where $A$ is a positive
constant and $B$ represents the predation rate of the anticancer cells.\\
As we do not intend to investigate the constant $A$, it was
concealed in the model for the convenience of discussion. The
transformation parameters can be given by \cite{3, 5, 7}:
\begin{equation}\label{2}
x=\frac{X}{A}, r=Ar_{B},q=\frac{K_{B}}{A},\tau=\frac{t}{A},\beta=B,
\end{equation}
and by substituting this into Eq. (1), we obtain:
\begin{equation}\label{3}
  \frac{dx}{d\tau}=rx\left(1-\frac{x}{q}\right)-\frac{\beta x^{2}}{1+x^{2}},
\end{equation}
where $r$ is the tumor cell growth rate and $\beta$ is the predation
rate of anticancer cells. By letting $\frac{dx}{d\tau}=0$, we can
obtain the steady states of the system from the Eq. (3). Clearly,
one of the solutions is $x=0$, while the other solutions satisfy:
\begin{equation}\label{4}
r\left(1-\frac{x}{q}\right)=\frac{\beta x}{1+x^{2}}.
\end{equation}
When $r$ varies and keeps $\beta$ and $q$ constant, the number of
solutions (namely equilibria) changes between one and
three\cite{32}. The range with three solutions changes with the
values of $\beta$ and $q$. This also occurs for a variable $\beta$
(or $q$)
and fixed $r$ and $q$ (or $\beta$).\\
Based on Eq. (3) (let $f(x)=\frac{dx}{d\tau}$), we can draw the
curves of $f(x) - x$ as shown in Fig. 1. For a curve with $r=1.0$
and $\beta=2.0$, $x=0$ and $x=x_{2}$ are unstable states, since
${\partial f}/{\partial x}>0$ at $x=0$, $x_{2}$. However, $x_{1}$
and $x_{3}$ are stable steady states since ${\partial f}/{\partial
x}<0$ at these two points.
 \begin{figure}[htbp]
   \begin{center}
   \caption{The equilibria of $f(x)$ vary with a decreases in $r$ and increases in $\beta$ for $q=10.0$ (arbitrary units).}\label{1}
\end{center}
\end{figure}\\
In Fig. 1, when $r$ decreases or $\beta$ increases, the solutions
$x_{2}$ and $x_{3}$ will disappear and only the stable state $x_{1}$
on the left side can be observed (e.g. the curve for $r=1.0$,
$\beta=3.0$). In contrast, when $r$ increases or $\beta$ decreases,
only the stable state $x_{3}$ on the right side can be observed (as
shown in Fig. 2). Clearly, $x_{1}$ is the refuge equilibrium, while
$x_{3}$ is the outbreak equilibrium. From a tumor control point of
view, we need to keep the number of tumor cells in the refuge state
rather than allowing it to reach an outbreak situation.\\
\begin{figure}[htbp]
   \begin{center}
  \caption{The equilibria of $f(x)$ vary with an increases in $r$ and decreases in $\beta$ for $q=10.0$ (arbitrary units).}\label{2}
\end{center}
\end{figure}
When considering the noise, the range of three equilibria is also
related to the noises. Similarly, the equilibrium state changes
between 1 and 3 by varying the parameters.
Therefore, it is necessary to investigate them by using the steady probability distribution function(SPDF).\\
\section{Tumor cell growth model with noise}
Equation (3) only describes the tumor growth behavior under ideal
conditions, without fluctuation. When considering a real situation,
external environmental disturbances such as the individual's state
of health, body temperature, and other disease may affect tumor
growth. In addition, artificial behavior like chemotherapy may also
have an effect. Because of these external disturbances, the growth
rate of tumor cells and the predation rate of anticancer cells may
vary greatly. In our work, the effects of additive noise on the
birth rate of tumor cell and multiplicative noise on the predation
rate of anticancer cells are further considered.
Hence the tumor growth equation considering external disturbances can be rewritten as: \\
\begin{equation}\label{5}
\frac{dx}{d\tau}=rx\left(1-\frac{x}{q}\right)-\frac{(\beta +\xi(t))
x^{2}}{1+x^{2}}+\Gamma(t),
\end{equation}
where $\xi(t)$ and $\Gamma(t)$ are Gaussian white noises. They have
the following properties:
\begin{equation}\label{6}
<\xi(t)\xi(t^{'})>=2\sigma\delta(t-t^{'}),
 \indent <\Gamma(t)\Gamma(t^{'})>=2D\delta(t-t^{'}),
 \end{equation}
where $\sigma$ and $D$ are the strength of the noises $\xi(t)$ and
$\Gamma(t)$, respectively. From this, we can derive the
corresponding Fokker-Planck equation for the evolution of SPDF based
on Eq. (5) and Eq. (6). The equation is as follows [11]:
\begin{equation}\label{7}
 \frac{\partial P(x,t)}{\partial
t}=-\frac{\partial A(x)P(x,t)}{\partial
x}+\frac{\partial^{2}B(x)P(x,t)}{\partial x^{2}},
\end{equation}
where $P(x,t)$ is the probability of the relative numbers of tumor
cells, and:
\begin{equation}\label{8}
 A(x)=rx\left(1-\frac{x}{q}\right)-\frac{\beta
x^{2}}{1+x^{2}}+\sigma\frac{2x^{3}}{(1+x^{2})^{3}},
\end{equation}
\begin{equation}\label{9}
  B(x)=\sigma\left[\frac{x^{2}}{1+x^{2}}\right]^{2}+D.
\end{equation}\\
\section{steady state analysis of the model}
Usually, what we are concerned with is the steady state. For Eq.
(7), when $\frac{\partial P(x,t) }{\partial t}=0$, we can obtain the
SPDF of the tumor cells  [11]:
\begin{equation}\label{10}
P_{st}(x)=\frac{N}{B(x)}\exp\left[\int^{x}\frac{A(x)}{B(x)}dx\right],
\end{equation}
 or
\begin{equation}\label{11}
P_{st}(x)=\frac{N}{B(x)}\exp[{M(x)}],
\end{equation}
where $N$ is the normalization constant,
\begin{equation}\label{12}
M(x)=\int^{x}\frac{A(x)}{B(x)}dx\ .
\end{equation}
When considering Eqs. (8), (9), and (12) together, we can obtain:
  \begin{eqnarray}\label{13}
M(x)&=&ax+bx^{2}+cx^{3}+d\ln|E(x)|+l\ln\left|\frac{U(x)}{V(x)}\right|+\ln\frac{\sqrt{E(x)}}{1+x^{2}}\nonumber\\
 &
&+m\arctan H(x)+n\left(\arctan K(x)+\arctan L(x)\right),
\end{eqnarray}
where
\begin{equation}\label{14}
a=-\frac{2r\sigma+q\beta(D+\sigma)}{q(D+\sigma)^{2}},
\end{equation}
\begin{equation}\label{15}
b=\frac{r}{2(D+\sigma)},
\end{equation}
\begin{equation}\label{16}
c=-\frac{r}{3q(D+\sigma)},
\end{equation}
\begin{equation}\label{17}
d=\frac{r\sigma}{2(D+\sigma)^{2}},
\end{equation}
\begin{equation}\label{18}
l=\frac{r\sigma\left(-3D+\sigma+2\sqrt{D(D+\sigma)}\right)+q\beta(D+\sigma)\left(\sigma-D+\sqrt{D(D+\sigma)}\right)}
{4q(D+\sigma)^{\frac{5}{2}}\sqrt{2\sqrt{D(D+\sigma)}-2D}},
\end{equation}
\begin{equation}\label{19}
m=\frac{r\sqrt{\sigma}(\sigma-D)}{2\sqrt{D}(D+\sigma)^{2}},
\end{equation}
\begin{equation}\label{20}
n=\frac{r\sigma\left(3D-\sigma+2\sqrt{D(D+\sigma)}\right)+q\beta(D+\sigma)\left(D-\sigma+\sqrt{D(D+\sigma)}\right)}
{2q(D+\sigma)^{\frac{5}{2}}\sqrt{2\sqrt{D(D+\sigma)}+2D}},
\end{equation}
\begin{equation}\label{21}
E(x)=D+2Dx^{2}+(D+\sigma)x^{4},
\end{equation}
\begin{equation}\label{22}
U(x)=\sqrt{D}+\sqrt{2\sqrt{D(D+\sigma)}-2D}x+\sqrt{(D+\sigma)}x^{2},
\end{equation}
\begin{equation}\label{23}
V(x)=\sqrt{D}-\sqrt{2\sqrt{D(D+\sigma)}-2D}x+\sqrt{(D+\sigma)}x^{2},
\end{equation}
\begin{equation}\label{24}
H(x)=\frac{(D+\sigma)x^{2}+D}{\sqrt{D\sigma}},
\end{equation}
\begin{equation}\label{25}
K(x)=\frac{2\sqrt{D+\sigma}x+\sqrt{2\sqrt{D(D+\sigma)}-2D}}{\sqrt{2\sqrt{D(D+\sigma)}+2D}},
\end{equation}
\begin{equation}\label{26}
L(x)=\frac{2\sqrt{D+\sigma}x-\sqrt{2\sqrt{D(D+\sigma)}-2D}}{\sqrt{2\sqrt{D(D+\sigma)}+2D}}.
\end{equation}\\
Equation (11) is the main result in this study. Based on this
equation, we can plot the figures and obtain the curves for
$P_{st}(x)$, $r$, $D$, $\beta$, and $\sigma$. In this way, we can determine the mechanisms of tumor growth model.\\
\section{Results and discussion}
\subsection{The relationships  between $r$, $\beta$, and $x$ under SPDF extremum condition}
In order to discuss the effects of the fluctuation on the steady
probability distribution (SPD) of tumor growth, it is necessary to
discuss the relationships between $r$, $\beta$, and $x$. The
condition to obtain the extremum of SPDF is:
 \begin{equation}\label{27}
 A(x)-B^{'}(x)=0,
 \end{equation}
By considering Eqs. (8), (9), and (27) together, we can obtain:
\begin{equation}\label{28}
 r\left(1-\frac{x}{q}\right)-\frac{\beta
x}{1+x^{2}}-\sigma\frac{2x^{2}}{(1+x^{2})^{3}}=0. \end{equation}
 Based on Eq. (28), we can draw the curves of $r$ - $x$ and $\beta$ - $x$ in Fig. 3 and 4, respectively.
As shown in Fig. 3, when $0.1 \leq \sigma \leq 1.0$, one value of
$r$ ($0.9 \leq r \leq 1.3$) corresponds to three values of $x$ with
a fixed $\beta$ and $q$. That is, the SPDF of tumor growth will
exhibit two peaks (corresponding to the three solutions, namely the
two stable steady states in the section 2) if $r$ is in that range.
For different values of $\sigma$ (or $\beta$), the curves almost
overlap with each other, except in the position around $x=1$. This
demonstrates that only when $x$ is around 1 can we observe the
difference of $r$ for different values of $\sigma$ (or $\beta$).
\begin{figure}[htbp]
   \begin{center}
  \caption{Plot of the SPDF extrema as a function of $x$ for different $\sigma$ values,
  using $\beta=2.25$ and $q=10.0$ (arbitrary units). }\label{3}
\end{center}
\end{figure}\\
\begin{figure}[htbp]
   \begin{center}
  \caption{Plot of the predation rate $\beta$ of anticancer cells as a function of $x$ for different values of $r$
  and $\sigma$ with $q=10.0$ under the SPDF extremum condition (arbitrary units).}\label{4}
\end{center}
\end{figure}\\
Similarly, adopting $r=1.0$ in Fig. 4, one value of $\beta$
($1.7\leq \beta \leq 2.5 $) corresponds to three values of $x$. The
range of the two peaks is different with different $r$ values.\\
As Eq. (28) is irrelevant to $D$, the ranges of the two peaks is
mainly determined by $r$ and $\beta$. In addition to the
aforementioned ranges, for smaller or larger values of $r$ (or
$\beta$), each curve has only one peak (mono-stability). This
corresponds to the situation shown in the section 2, but there are
four parameters here($r$, $\beta$, $q$ and $\sigma$).\\
It is clear that the SPD of tumor growth switches between
bi-stability and mono-stability when the parameter $r$ changes. As
shown in Fig. 5, the positions of the peaks vary with the different
$r$ values. For a small $r$ value, there is only one peak on the
left side, which represents a small quantity of tumor cells in
healthy people or the annihilation of tumor cells. With the increase
of $r$ value, two peaks can be observed. With a further increase in
the $r$ value, the peak number is again reduced to one and its
position shifts to the right side, indicating the steady growth of
the tumor cells. Comparing the growth behaviors under different $r$
values, it can be concluded that a large value of $r$ is favorable
for tumor cell growth.\\
\begin{figure}[htbp]
   \begin{center}
  \caption{Plot of $P_{st}(x)$ against $x$ for
 different $r$ values, using $\beta=2.3, q=10.0, D=0.5$, and $\sigma=0.5$ (arbitrary units).}\label{1}
\end{center}
\end{figure}\\
\begin{figure}[htbp]
   \begin{center}
  \caption{Plot of $P_{st}(x)$ against $x$ for
 different $\beta$ values, using $q=10.0, r=1.0, D=0.5$, and $\sigma=0.5$ (arbitrary units).}\label{1}
\end{center}
\end{figure}\\
Fig. 6 shows the effects of the $\beta$ value on the tumor growth.
With the increase of the $\beta$ value, the number of peaks on the
curve changes from one to two, and finally
back to one. It is clear that a large predation rate is unfavorable for tumor growth. \\
\subsection{The effects of fluctuations on tumor growth rate}
Figure 7 shows the effects of $D$ on tumor growth for a small
$\beta$ value (e.g. $\beta=1.7$). When the value of $D$ increases,
the peak intensity of the probability density decreases, which
demonstrates
that an increase in $D$ is unfavorable for tumor growth.\\
\begin{figure}[htbp]
   \begin{center}
  \caption{Plot of $P_{st}(x)$ against $x$ for
 different $D$ and $\sigma$ values, using $\beta=1.7, q=10.0$, and $r=1.0$ (arbitrary units).}\label{1}
\end{center}
\end{figure}
In the case of a large $\beta$ value (e.g. $\beta=2.6$), the peak
position shifts to the left side, as shown in Fig. 8. When the $D$
value increases, the SPD of tumor cells moves to the more positive
$x$ direction. When healthy people receive chemotherapy, normal
tissue cells and anticancer cells may be killed rather than tumor
cells. Therefore, in such a case, chemotherapy is favorable for
tumor growth (although there is usually a small quantity of tumor
cells
in healthy people, they can be controlled by the immunity of the human body).\\
It can thus be concluded that $\beta$ plays a very important role in
tumor growth, especially if $\beta$ is very small or very large,
while $\sigma$ is less important for the tumor growth. In a real
situation, the magnitude of the predation rate can determine the body's anticancer ability.\\
\begin{figure}[htbp]
   \begin{center}
  \caption{Plot of $P_{st}(x)$ against $x$ for
 different $D$ and $\sigma$ values, using $\beta=2.6, q=10.0$, and $r=1.0$ (arbitrary units).}\label{8}
\end{center}
\end{figure}
For a moderate $\beta$ value, for example $\beta=2.26$, there are
two peaks for all the $D$ values, as shown in Fig. 9 and 10. For a
fixed $\beta$, the SPDF curve becomes flatter with an increase in
$D$. The noise interferes with the tumor growth as well as the
predation rate of anticancer cells. For example, by using
chemotherapy, tumor cells may be extinguished. At the same time,
normal tissue cells and anticancer cells may also be damaged. Here,
whether or not the increase of $D$ is favorable for tumor growth is
determined by $\sigma$. For a small value of $\sigma$, as shown in
Fig. 9 ($\sigma=0.1$), the intensity of the right peak decreases
with an increase in $D$, which means that it is unfavorable for
tumor growth. In contrast, for a large value of $\sigma$, as shown
in Fig. 10 ($\sigma=0.8)$, even though the intensities of two peaks
decrease at the same time, the SPD of tumor cells moves to the more
positive direction of $x$. That is, when $\sigma$ is larger,
interference with the anticancer cells is dominant. Thus, the
increase of $D$ in the case of a larger $\sigma$
when $\beta$ is a middle value is favorable for tumor growth.\\
\begin{figure}[htbp]
   \begin{center}
  \caption{Plot of $P_{st}(x)$ against $x$ for
 different $D$ values, using $\beta=2.26, q=10.0, r=1.0$, and $\sigma=0.1$ (arbitrary units).}\label{9}
\end{center}
\end{figure}
\begin{figure}[htbp]
   \begin{center}
  \caption{Plot of $P_{st}(x)$ against $x$ for
 different $D$ values, using $\beta=2.26, q=10.0, r=1.0$, and $\sigma=0.8$ (arbitrary units).}\label{10}
\end{center}
\end{figure}
For a different value of $r$, for example, $r=1.2$, the results
indicate that the laws acting on $P_{st}(x)$ and $D$ are same as
that of $r=1.0$. However, the range of the two peaks is different.
If the curves for $r=1.0$ shift to the more positive direction of
$\beta$, they can nearly overlap with the curves for $r=1.2$. For
example, Fig. 11 is almost same to Fig. 9, except that the relevant
$\beta$ value (2.72) is higher by around 0.46 larger than that for
$r=1.0$ ($\beta=2.26$). This means that the laws of tumor growth are
similar even with a different tumor growth rate.\\
\begin{figure}[htbp]
   \begin{center}
  \caption{Plot of $P_{st}(x)$ against $x$ for
 different $D$ values, using $\beta=2.72, q=10.0, r=1.2, $ and $\sigma=0.1$ (arbitrary units).}\label{11}
\end{center}
\end{figure}
\subsection{ The effects of fluctuations on predation rate }
Different values of $r$, $\beta$, $D$, and $\sigma$ are adopted in
this study. The corresponding results show that the effects of
fluctuations on $\beta$ are similar to those on $r$. For a small
value of $\beta$, an increase in $\sigma$ is unfavorable for tumor
growth. For a large value of $\beta$, on the other hand, an increase
in $\sigma$ is favorable for tumor growth. This effect is
independent of $D$. However, for a moderate value of $\beta$,
whether or not the noise is favorable for tumor growth  is
determined by $D$. For a small value of $D$, interference with
$\beta$ is dominant; this is favorable for tumor growth. In
contrast, large values of $D$ are unfavorable for tumor growth. If
$r$ is changed, the laws are same as those mentioned above, but the
corresponding $\beta$ will be
different. This is also similar to the effects of the fluctuation of $r$.\\
\section{summary}
We investigated the effects of the environmental disturbances on
tumor cell growth. By solving the corresponding Fokker-Planck
equation, we obtained analytic expressions of the steady state
probability distribution of tumor cells. It was found that the
effects of noise on the tumor growth are mainly determined by the
predation rate $\beta$: (1) For a small value of $\beta$, the
effects of the disturbance on tumor growth and anticancer cells are
unfavorable for tumor growth; (2) A large value of $\beta$ is
favorable for tumor growth. (3) For a moderate value of $\beta$, the
effect is determined by the fluctuation in the relative strength of
the two noises: (a) If the fluctuation strength of the predation
rate $\sigma$ is small, the increase of the tumor growth rate
fluctuation intensity $D$ is unfavorable for tumor growth; in
contrast, it is favorable for tumor growth if $\sigma$ is large; (b)
if $D$ is small, increasing $\sigma$ is favorable for the tumor
growth; in contrast, decreasing $\sigma$ is unfavorable for tumor
growth. Although further work is still necessary, it is believed
that the present results can give some useful insights
for the clinical treatment of tumors. \\

This work was supported by National Natural Science Foundation of
China under Grant No. 30600122.\\

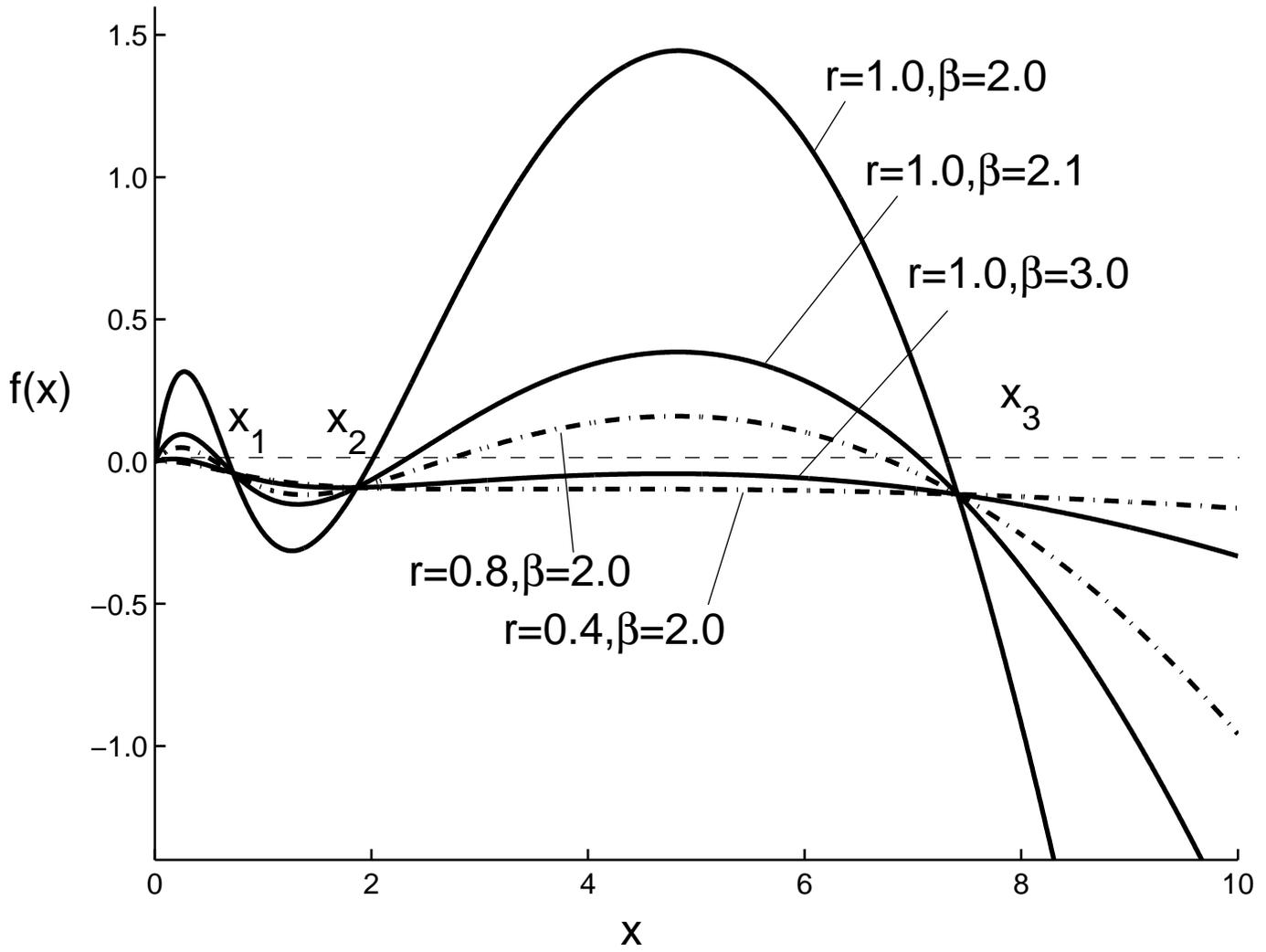
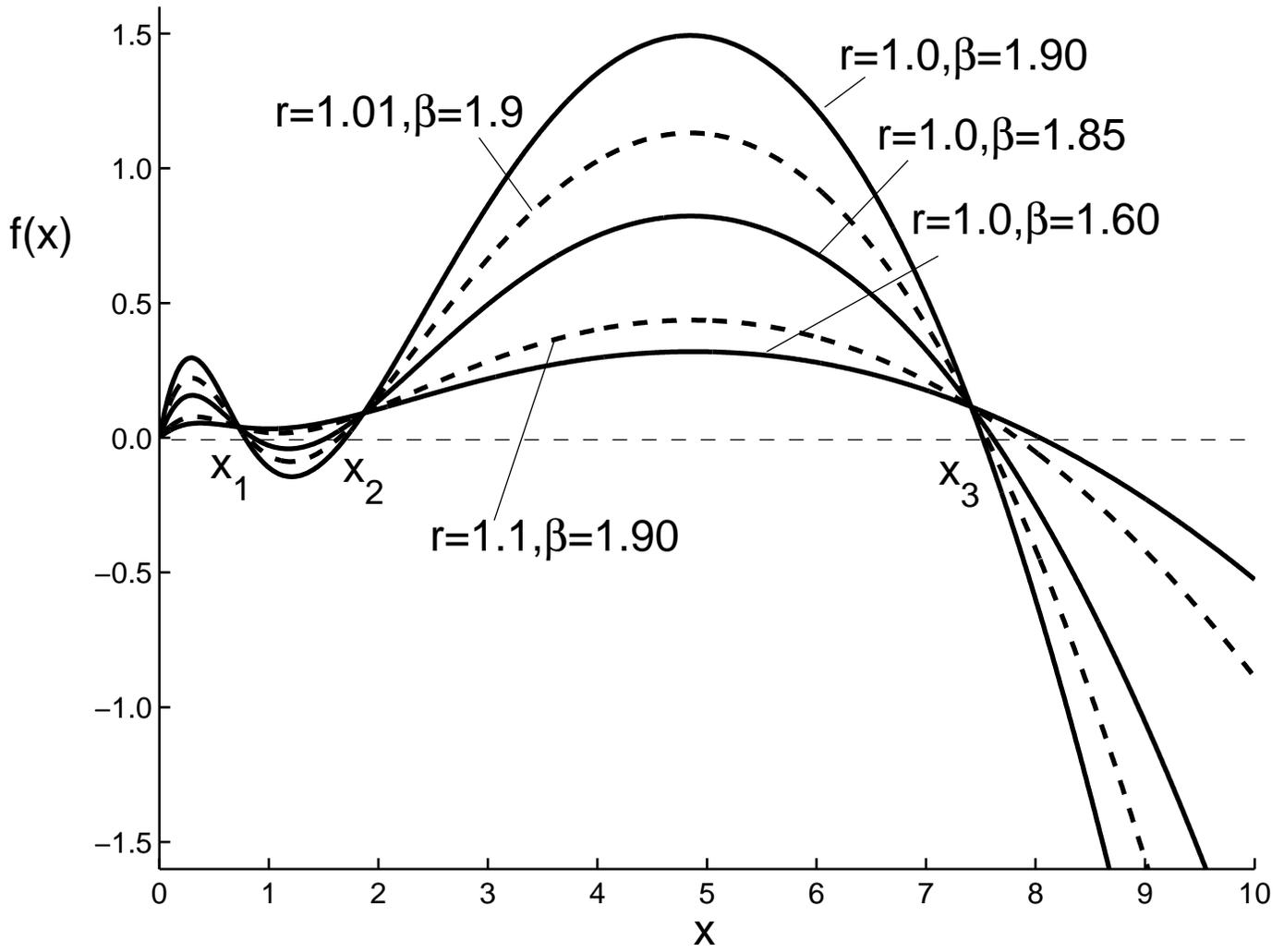
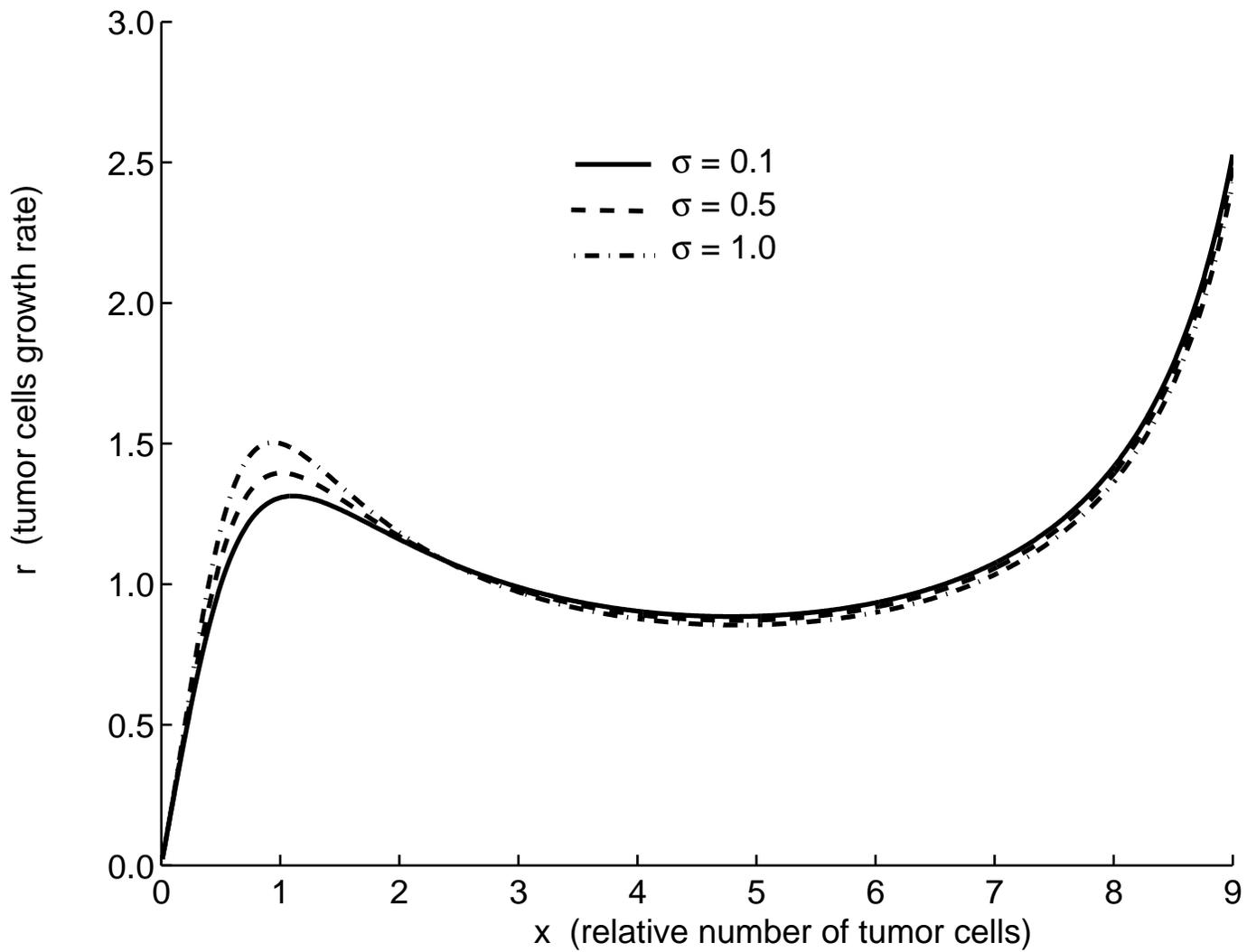
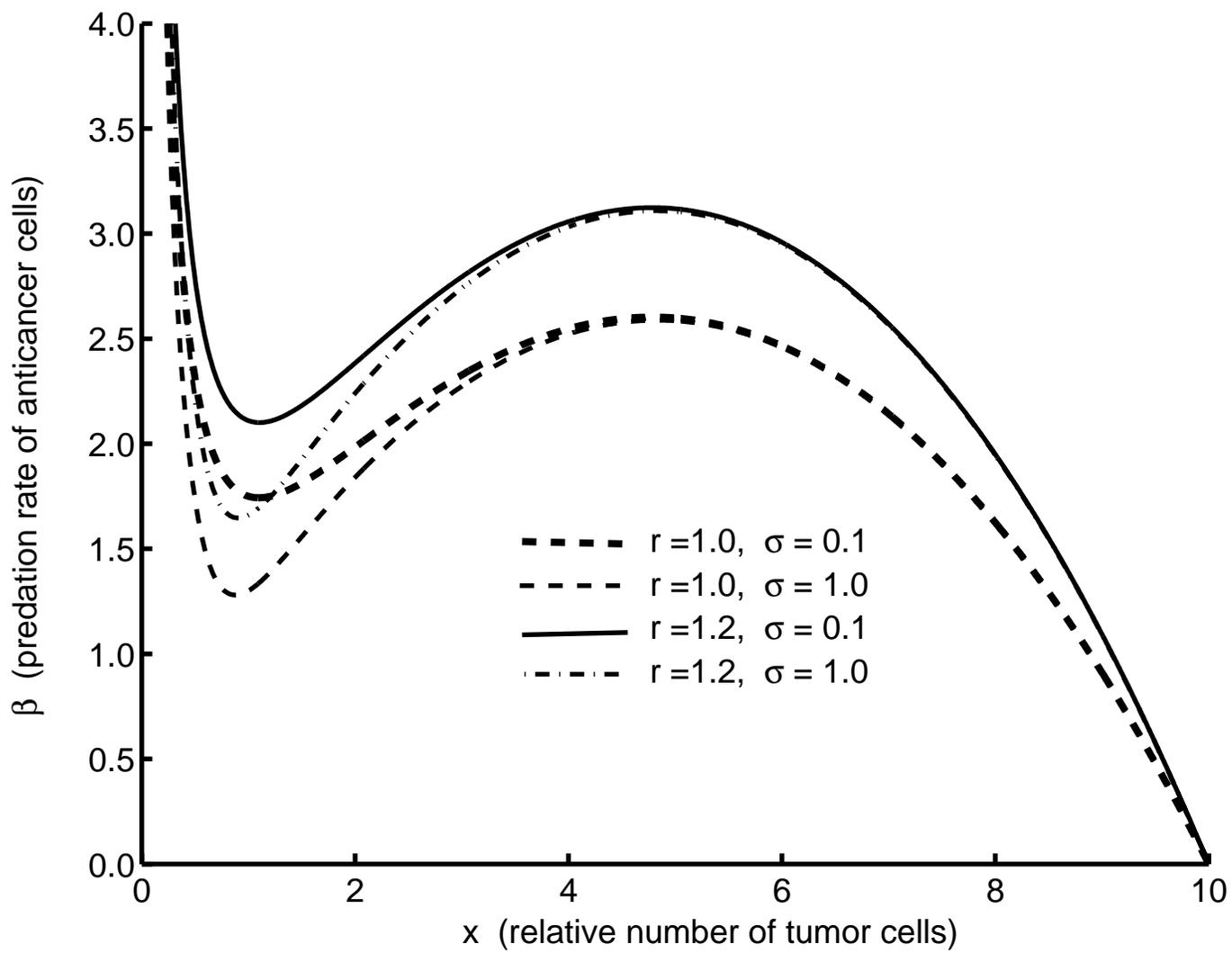
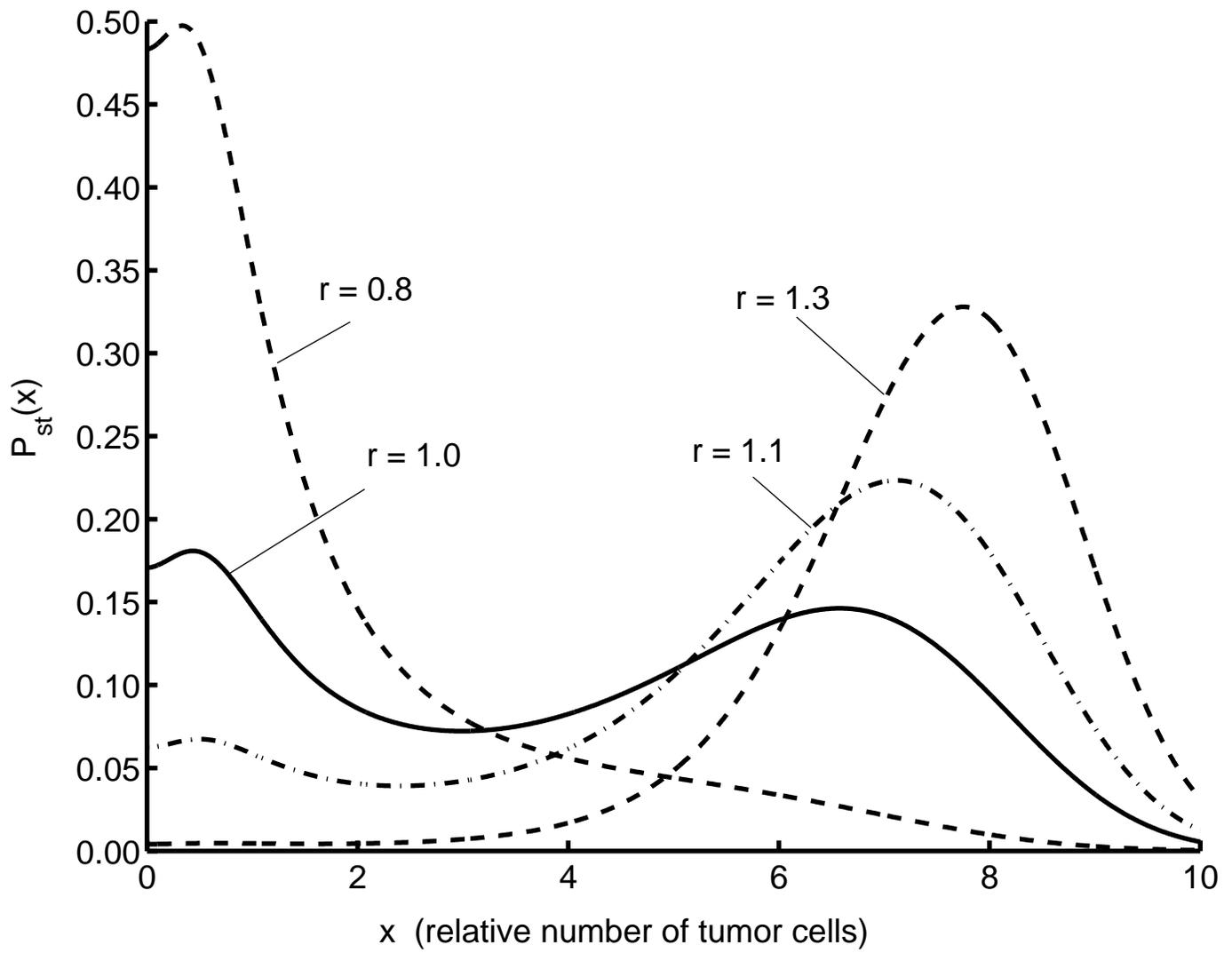
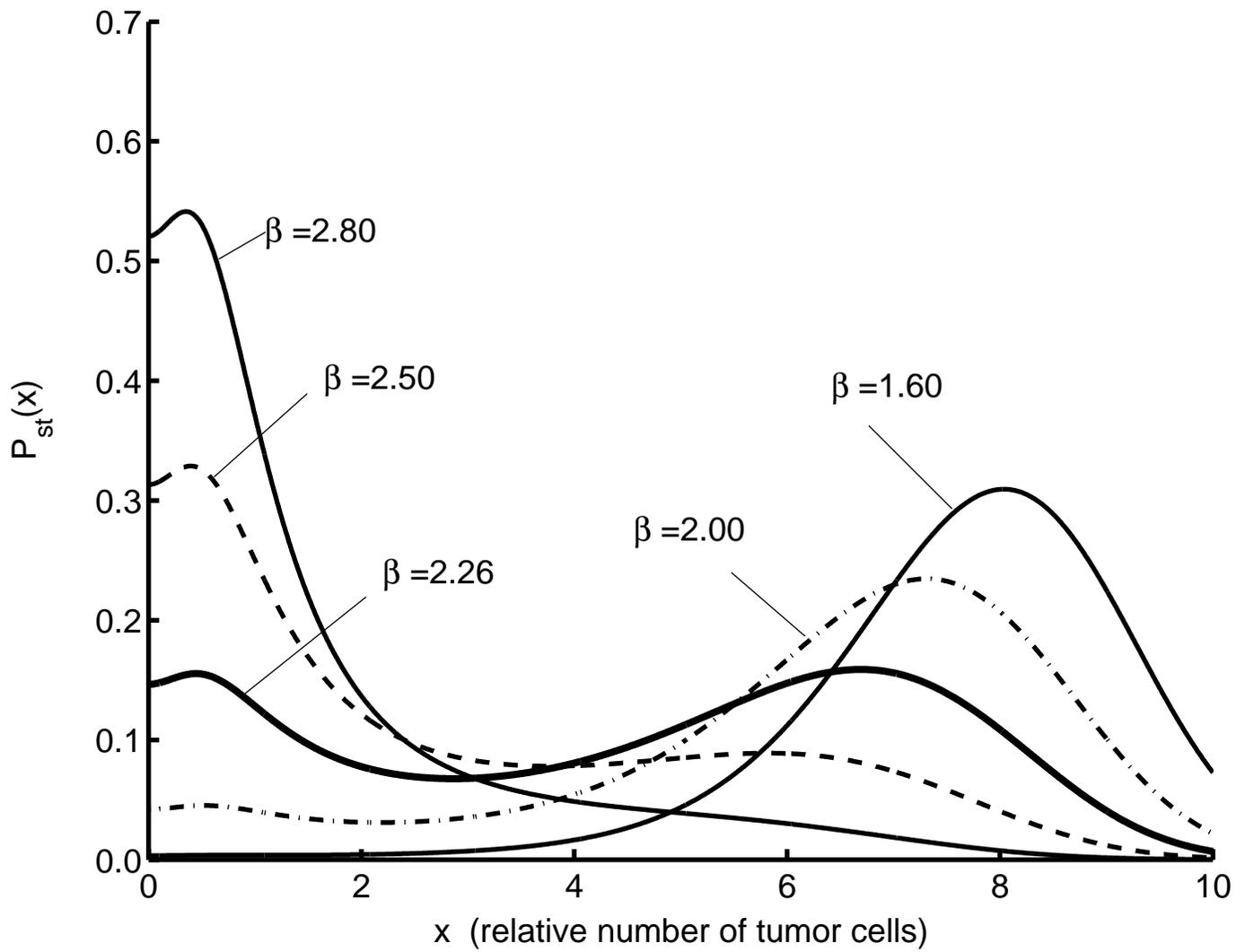
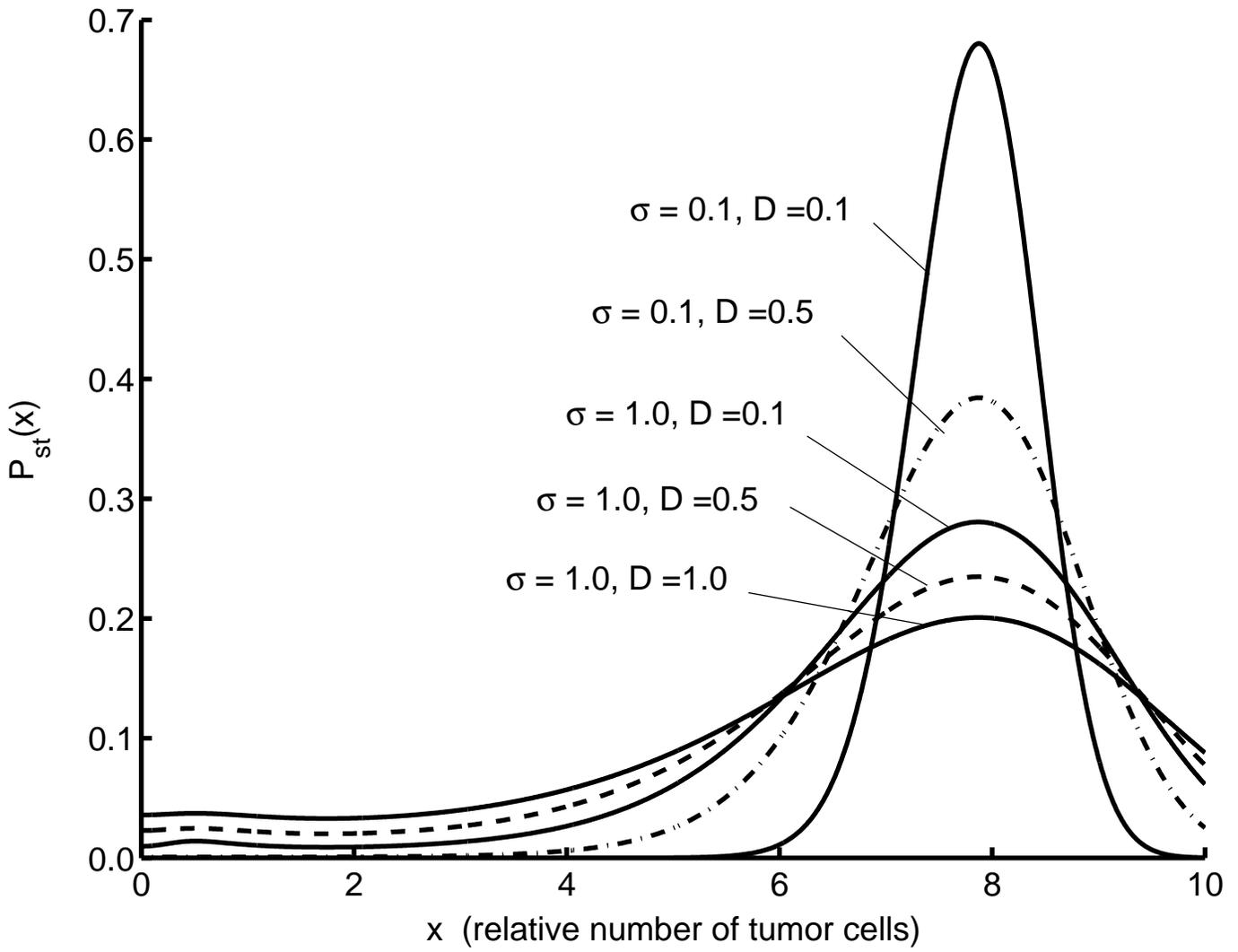
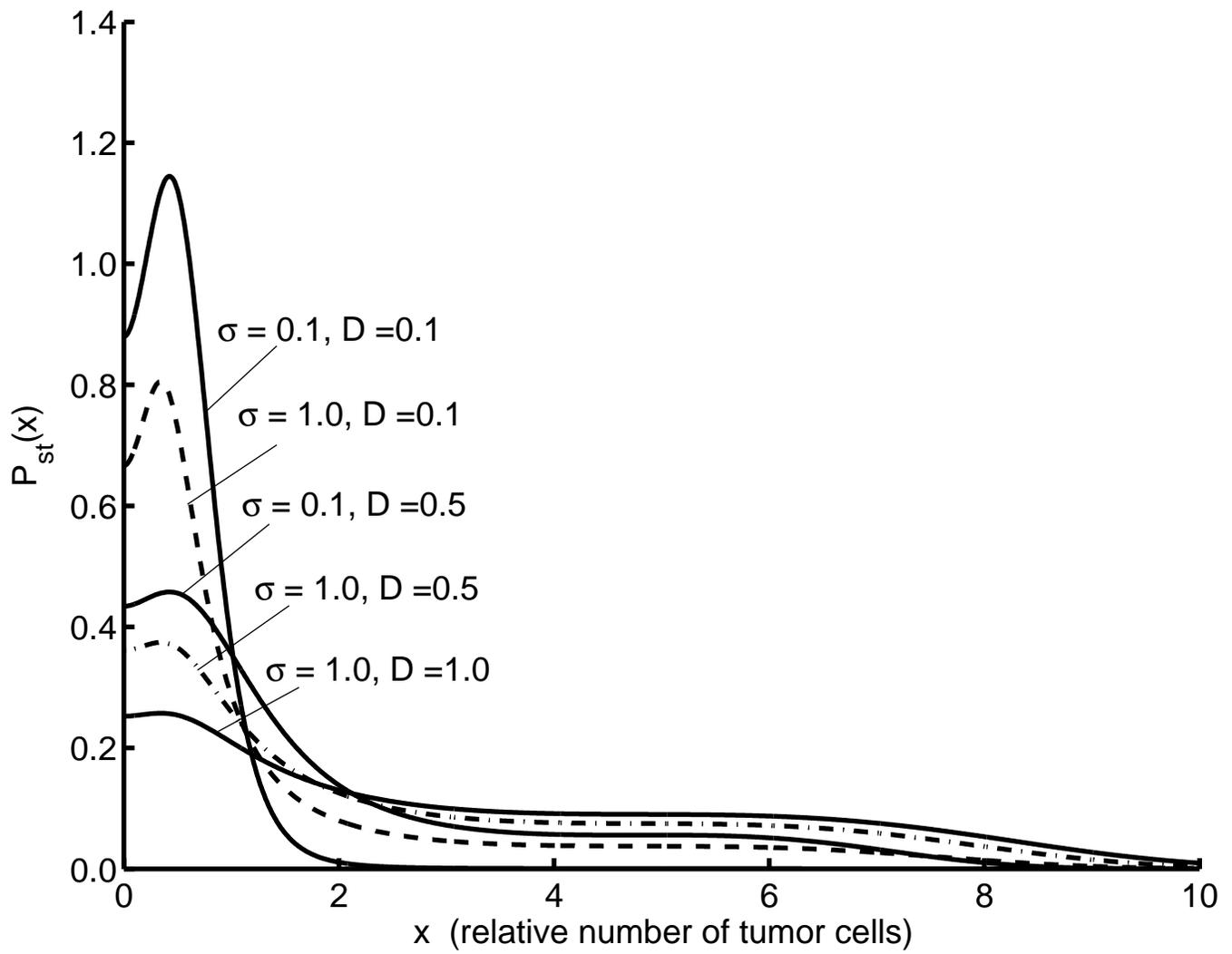
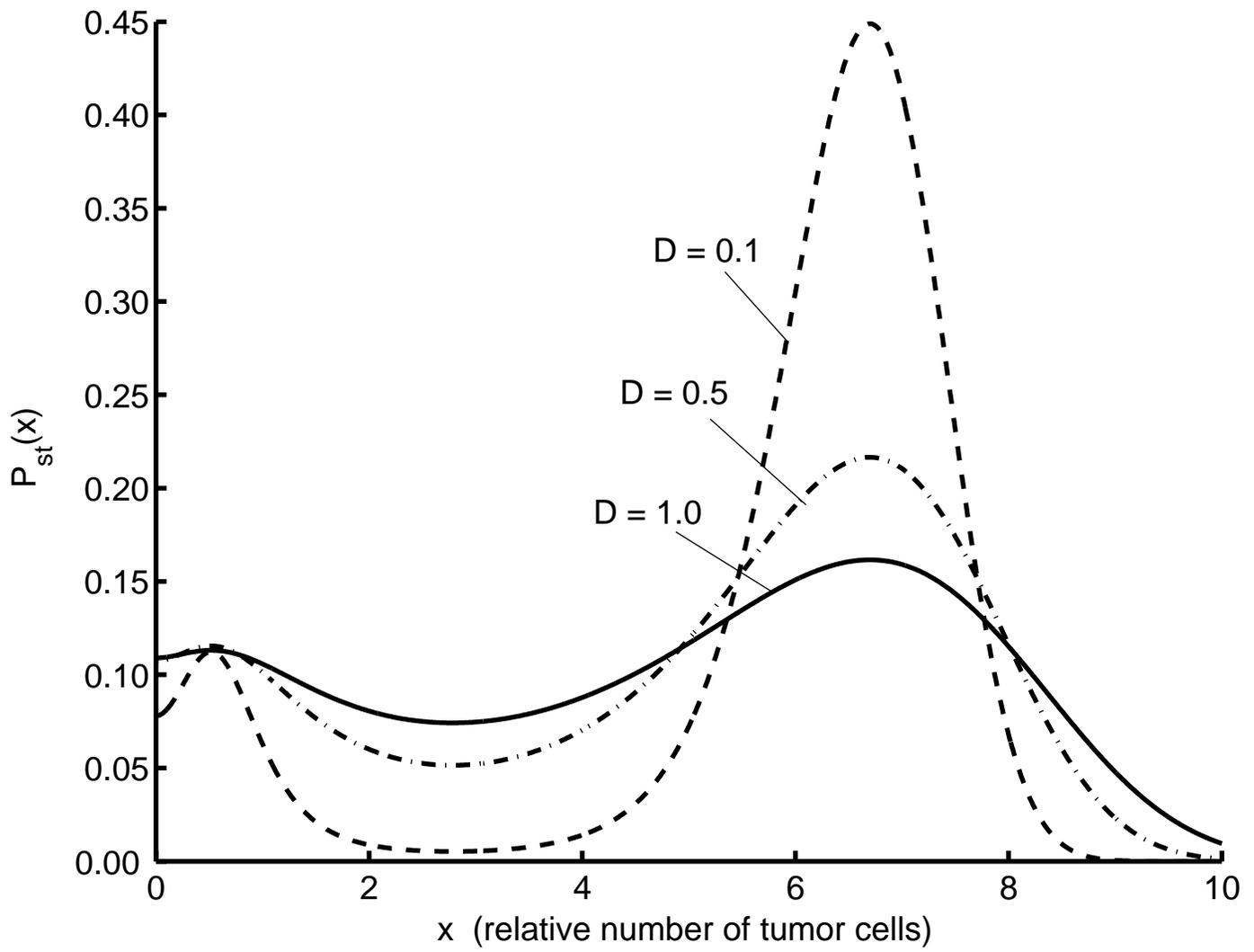
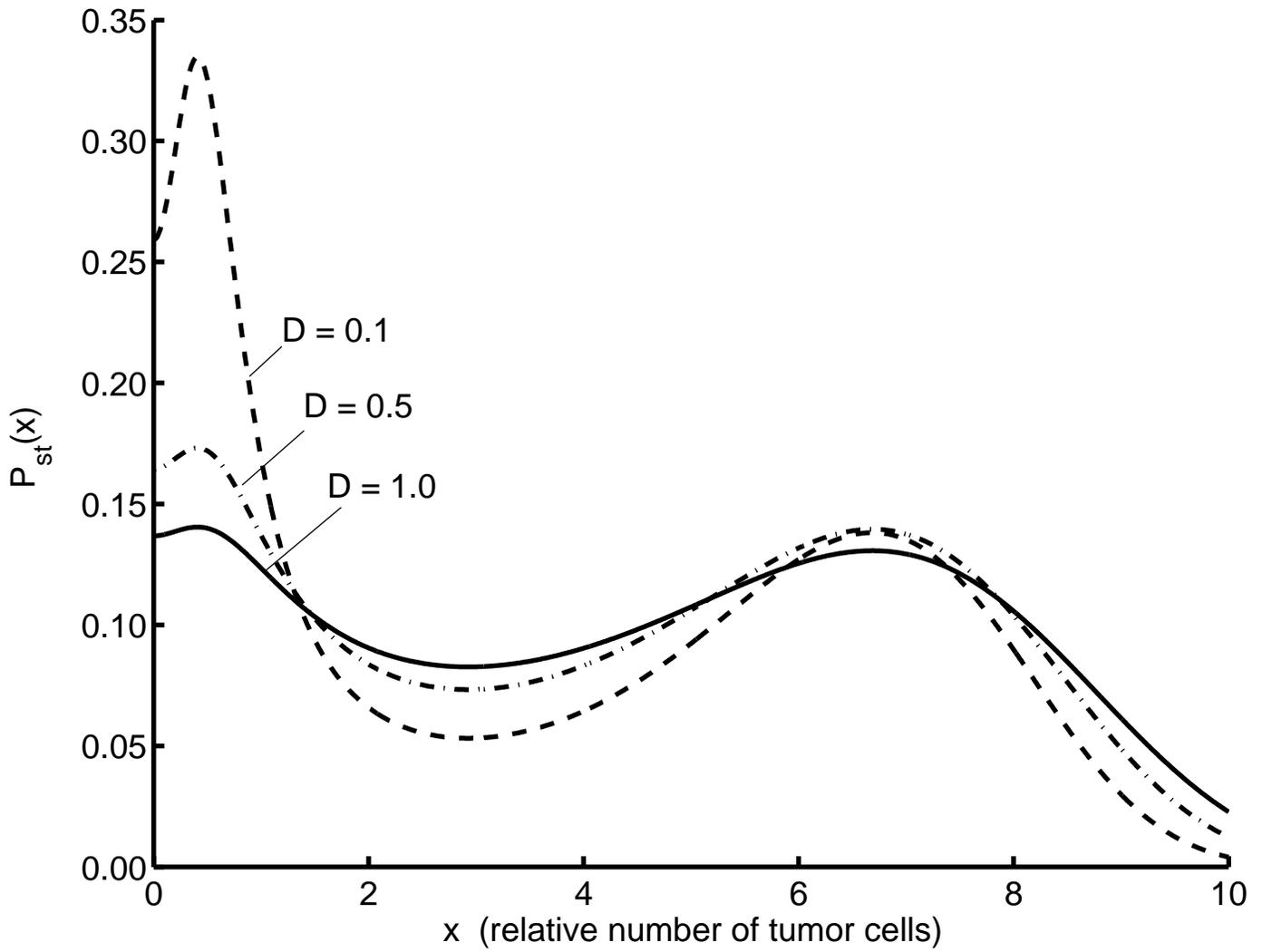
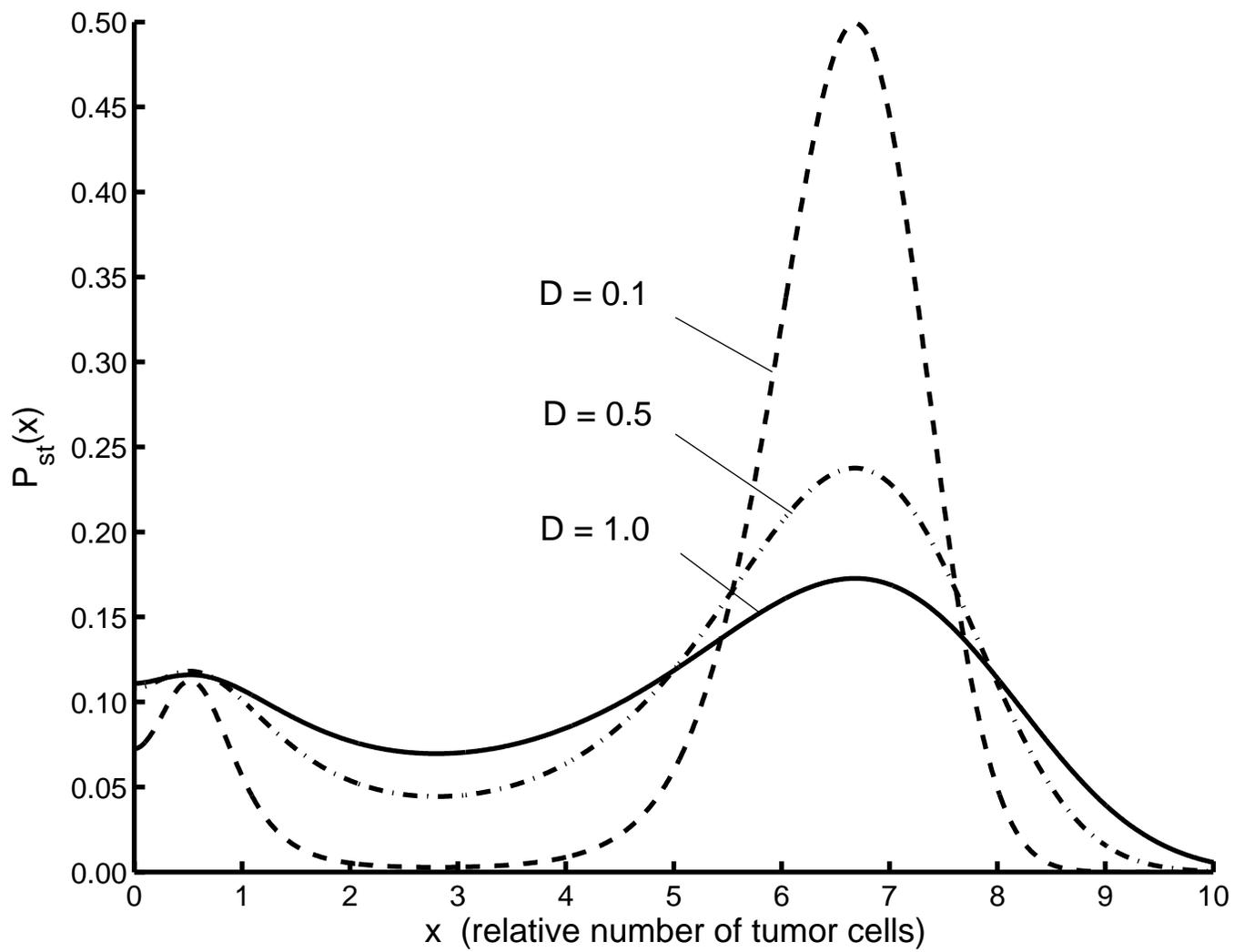
\end{document}